\documentclass[prb,amssymb,twocolumn,superscriptaddress,showpacs]{revtex4}
\usepackage{bm}
\usepackage{graphicx}
\usepackage{color}
\usepackage{amsmath}

\newcommand{\mr}[1]{{{\mathrm{#1}}}}
\newcommand{\w}{\omega}

\newcommand{\aid}{a^{\dagger}_{i}}
\newcommand{\ai}{a^{\phantom{\dagger}}_{i}}

\begin{document}

\title{Optimal broadening of finite energy spectra in the numerical renormalization group: 
application to dissipative dynamics in two-level systems}

\author{Axel Freyn}
\author{Serge Florens}
\affiliation{Institut N\'eel, CNRS and Universit\'e Joseph Fourier, 25 avenue des Martyrs, BP 166, 38042 Grenoble, France}

\begin{abstract} 

Numerical renormalization group (NRG) calculations of quantum impurity models,
based on a logarithmic discretization in energy of electronic or bosonic
Hamiltonians, provide a powerful tool to describe physics involving widely
separated energy scales, as typically encountered in nanostructures and strongly
correlated materials. 
% This main advantage of the NRG over other numerical simulation schemes based on 
% linear discretization proved so far to be a drawback for resolving sharp spectral 
% features at finite energy, such as Hubbard band. 
This main advantage of the NRG was however considered a drawback for resolving sharp 
spectral features at finite energy, such as dissipative atomic peaks.
% We surprisingly find that the NRG spectra automatically encode all the necessary 
% information near dissipative resonances, however narrow these may be.
Surprisingly, we find a bunching of many-body levels in NRG spectra near dissipative 
resonances, and exploit this by combining the widely-used Oliveira's $z$-trick, 
using an averaging over {\it few} discrete NRG spectra, with an
optimized {\it frequency-dependent} broadening parameter $b(\w)$.
%self-consistently determined to obtain smooth and precise spectral functions.
%This strategy offers a gain in computational power of possibly several orders of magnitude 
%for a given accuracy as compared to $z$-averaging alone, and extracts all the
This strategy offers a tremendous gain in computational power and extracts all the
needed information from the raw NRG data without {\it a priori} knowledge of the
various energy scales at play. 
As an application we investigate with high precision the crossover from coherent to 
incoherent dynamics in the spin boson model.
%Other possible uses of the method will cover quantitative calculations in the dynamical 
%mean field theory.
\end{abstract} 
\maketitle

A general hallmark of many-particle interaction, as found in a variety of condensed 
matter systems such as nanostructures and strongly correlated materials, lies in 
the presence of several energy scales, possibly widely separated from each other 
due to renormalization and dynamical effects.
Two well-studied examples found in electronic systems concern the Kondo
effect for magnetic impurities in metals and Fermi liquids in proximity to a Mott 
insulating phase, two instances where low-energy quasiparticles emerge below a typical
temperature which is quite reduced from the bare Fermi energy~\cite{Hewson}.
These low-lying excitations do however coexist with higher energy atomic levels,
also broadened and displaced in a strong manner from their bare atomistic
values due to the dissipation brought by the electronic environment.
Such complex physical effects, taking place on a broad range of energies, entail 
great practical difficulties for most direct numerical approaches. These are partially 
lifted using Wilson's original idea of the logarithmic discretization~\cite{Wilson,Krishnamurthy}, as
implemented in numerical renormalization group calculations (see Ref.~\onlinecite{Bulla1}
for a recent review). This technique has been improved in the last twenty years
to calculate static and dynamic quantities both for fermionic~\cite{Costi1} and
bosonic quantum impurity models~\cite{Tong}. Important practical applications 
until now involve the calculation of transport in the Kondo regime for Kondo alloys
and artificial quantum dots~\cite{Costi2}, as well as the accurate description of the 
zero-temperature Mott transition by combining NRG~\cite{Bulla2} with Dynamical Mean Field
Theory (DMFT)~\cite{DMFT}. More generally, exponentially small energy scales are
also found in the vicinity of quantum critical points~\cite{Sachdev}, so that impurity 
models provide a simplified testbed for the theory of quantum critical phenomena~\cite{Vojta}. 
Again the NRG is the ideal technique for studying such impurity quantum phase 
transitions~\cite{Bulla1}, with potential implications for artificial 
nanostructures~\cite{Potok,Roch}.

Despite these successes, the foundation of NRG on a logarithmic discretization
mesh implies a loss in accuracy for the high energy spectral features, which has
plagued most calculations so far. Not only are atomic-like excitations
physically observable, using photoemission or tunneling spectroscopies,
but they may also be intimately tied to the low-energy excitations.
Such interesting behavior occurs in the vicinity of the Mott 
metal-insulator transition~\cite{DMFT}, where the self-consistently determined
electronic environment in the DMFT picture shows electronic states 
violently redistributed over all energy scales. The numerical cost of converging the 
DMFT equations certainly requires efficient and accurate NRG calculations for the 
spectral functions, without {\it a priori} knowledge of the excitations
involved. For this reason, the idea of averaging over $N_z$ realization of the
Wilson chain, using the so-called $z$-trick~\cite{Oliveira1} to fill in the missing
spectral information (to be discussed later on), seems prohibitive for 
most practical calculations, and offers a limited gain for very narrow spectral
structures (see Ref.~\onlinecite{Zitko} for a detailed study).

In this Rapid Communication, we show that the broadening procedure used to smoothen the 
discrete NRG data is one critical step for obtaining an optimal resolution at all 
energies. We henceforth develop a simple adaptive procedure where the standard 
broadening parameter $b$ is taken to be frequency dependent. This choice is dictated by our
surprising observation that the density of many-body NRG levels increases sharply as 
soon as narrow atomic resonances are encountered. Together with the usual $z$-trick, 
combining several but reasonably {\it few} NRG calculations, this extra $b(\w)$-trick 
allows to compute finite frequency spectra using as little as $N_z=10$ NRG calculations, 
in situations where large scale NRG $z$-averaging would be prohibitive.
%providing accuracies comparable to the standard averaging of $N_z\gg10$ NRG runs with a fixed
%broadening parameter $b\propto1/N_z$ that scales to zero with the resonance width. 
%As this may involve from several dozens to thousands of parallel NRG calculations, 
%depending on the regimes and parameters investigated, 
An improved broadening procedure leading to errors in spectral functions limited
to few percents could constitute a further step in applying the NRG to a wider class of 
problems, such as DMFT+NRG calculations.
%For most practical cases using such optimized $b(\w)$ and low values of $N_z$, errors 
%in spectral functions can easily be limited from fractions up to few percents,
%depending whether one probes at or away from resonances, meeting the necessary requirements for 
%controlled DMFT+NRG calculations. 

In order to explicitly demonstrate these ideas, we focus on the simplest quantum
impurity model, namely the spin boson Hamiltonian:
\begin{equation}
H = \frac{\Delta}{2}\,\sigma^x + \frac{\lambda}{2}\,\sigma^z \sum_i (\aid+\ai) +
\sum_i \w_i \aid \ai\, ,
\label{eq:ham}
\end{equation}
that involves a two-level system, described by a quantum spin $1/2$, and a bosonic 
bath $\aid$ with continuous spectrum $\w_i$ of energies.
$\Delta$ is a transverse magnetic field, while the coupling constant $\lambda$ controls the 
strength of longitudinal dissipation. The bosonic spectrum is assumed to be sub-ohmic
with bath exponent $0<s\leq 1$ (this includes the well-studied ohmic
case $s=1$):
\begin{equation}
J(\w) \equiv \sum_i \pi \lambda^2 \delta(\w-\w_i) = 2 \pi \alpha
\w_c^{1-s}\w^s\theta(\w)\theta(\w_c-\w)
\label{eq:J}
\end{equation}
where $\w_c$ is a high-energy cutoff.
For small values of the dissipation $\alpha$, this model is known to exhibit coherent precession 
of the spin around the $x$ axis at zero temperature. By increasing the coupling
to the bath for values of $s\lesssim1$, Rabi oscillations are progressively damped~\cite{Leggett}, 
before the two-level system localizes in one potential minima via a quantum phase
transition~\cite{Tong}. When $s\gtrsim0$ however, the phenomena of localization and 
decoherence occur in reverse order~\cite{Anders}. 
Despite its simplicity, this model embodies all the effects typical of strong
correlations: low-energy scales are indeed dynamically generated near the localization
transition, while discrete spin excitations due to the transverse field $\Delta$ are
deeply affected by the bosonic environment, that leads to broadening and frequency
shift in the magnetic response \begin{equation}
C(\w)=\frac{1}{2\pi} \int_{-\infty}^{+\infty} 
\!\!\!\!{\rm d}t \, C(t) e^{i \w t}
\end{equation}
with $C(t) = \frac{1}{2}\langle [\sigma_z(t),\sigma_z] \rangle$ the spin
autocorrelation function.

\begin{figure}[ht]
\includegraphics[width=8.5cm]{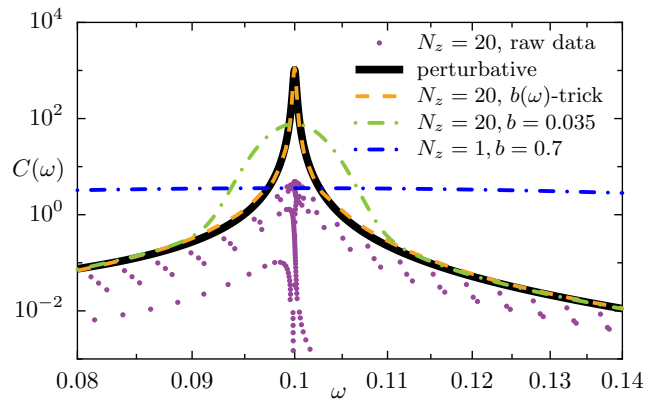}
\caption{(Color online) Spin susceptibility $C(\w)$ of the sub-ohmic spin boson model at
$s=0.1$, $\Delta=0.1\w_c$ and $\alpha=0.000125$. Raw data
$| \langle 0,z | \sigma_z | a,z \rangle |^2/[2N_z (\epsilon_{a,z}-\epsilon_{0,z})]$ are 
given as circles for $N_z=20$ combined NRG calculations, solid line is the perturbative result~(\ref{eq:analytic}), 
and the three dashed lines are the various NRG broadenings discussed in the text.} 
\label{fig:raw}
\end{figure}

The implementation of the NRG procedure follows the standard route~\cite{Wilson}
as discussed in Ref.~\onlinecite{Tong} for the extension to bosonic Hamiltonians. The
bosonic band~(\ref{eq:J}) is logarithmically discretized with the
Wilson parameter $\Lambda>1$, first on the highest energy interval 
$[\Lambda^{-z}\w_c,\w_c]$, and then iteratively on successive decreasing energy
windows $[\Lambda^{-n-z}\w_c,\Lambda^{-n-z+1}\w_c]$ for $n$ strictly positive integer.
This also introduces the crucial $z$ parameter ($0<z<1$) that is used to average
over $N_z$ Wilson chains~\cite{Oliveira1}, allowing to obtain better resolution on the 
finite energy states.
The rest of the NRG follows Ref.~\onlinecite{Tong}, coupling iteratively the kept
energy levels up to iteration $n$ to the states living in the shell
$n+1$, and truncating the successive Hamiltonians to keep up with a manageable
number of eigenstates. All subsequent calculations were performed with
$\Lambda=2$, $N_b=8$ kept bosonic states on the added bosonic ``site'', and $N=160$
kept NRG states, ensuring good convergence. The resulting discrete spectra
at successive NRG iterations are combined using the interpolation
scheme proposed in Ref.~\onlinecite{Bulla3} (see however Ref.~\onlinecite{Weichselbaum} 
for a more rigorous implementation), leading to a set of $z$-dependent many-body 
energy levels $\epsilon_{a,z}$ labelled by quantum number $a$.
The spin-spin correlation function is thus readily obtained at zero temperature
as:
\begin{equation}
C(\w )=\frac{1}{2 N_z}\sum_{a,z} | \langle 0,z | \sigma_z | a,z \rangle |^2
               \delta \left( |\w|+\epsilon_{0,z}-\epsilon_{a,z} \right)
\label{eq:discrete}
\end{equation}
where $\epsilon_{0,z}$ is the ground state energy, and these raw NRG data are displayed 
on Fig.~\ref{fig:raw}, to be discussed below.

Let us now discuss the general structure of the NRG spectra. For a single NRG
calculation with a given value of $z$, the {\it a priori} energy resolution at the scale
$\w_N=\Lambda^{-N-z}$ is $\delta\w_N=
\Lambda^{-N-z}-\Lambda^{-N-z-1}=(1-\Lambda^{-1}) \w_N$, so that the resolution 
degrades at increasing energy. For obtaining smooth NRG spectra, the delta-peaks
in~(\ref{eq:discrete}) are usually broadened~\cite{Bulla1} at energy $\w_N$ {\it on the same 
scale}:
\begin{equation}
\delta \left( |\w|-\w_N\right) \rightarrow
\frac{e^{-b^2/4}}{\w_N b \sqrt{\pi}}
e^{-[\log(|\w|/\w_N)^2/b]^2}
\label{eq:broaden}
\end{equation}
with $b\approx0.7$ typically.
Combining $N_z$ NRG runs using the $z$-averaging in~(\ref{fig:raw}) allows in
principle to improve by a factor $N_z$ the accuracy at high energy, since  the broadening parameter 
may now be decreased down to $b=0.7/N_z$. This procedure actually faces two problems: i)
for very sharp resonances (typically several orders of magnitude narrower that the
natural high-energy cutoff), parallelizing $N_z\gg10$ NRG calculations becomes
too prohibitive, especially with the aim of DMFT simulations; ii) states far from 
the resonances can then be too much underbroadened, so that oscillations of 
period $\Lambda/N_z$ can become quite prominent. 
In this view, quantitative spectra can be accurately extracted from the NRG data 
only in the continuum limit, either with $\Lambda\rightarrow1$ or 
$N_z\rightarrow+\infty$, as assumed in the previous
literature~\cite{Zitko,Oliveira2,Weichselbaum2}.
\begin{figure}[ht]
\includegraphics[width=8.5cm]{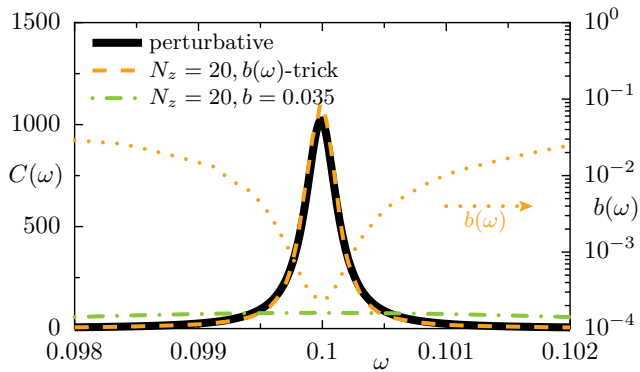}
\caption{(Color online) Comparison of various broadening procedures near the
resonance for the parameters of Fig.~\ref{fig:raw}: standard interleaved 
averaging with $N_z=20$ and $b=0.7/N_z=0.035$ (dash-dotted line), $b(\w)$-trick 
with $b_0=0.12$ (dashed line), and analytical result (full line). The 
adaptive broadening parameter $b(\w)$ is also given (note the 
logarithmic scale on the right).}
\label{fig:compare_verysharp}
\end{figure}

We wish to show however that NRG provides much more information close to
dissipative resonances than usually expected. For this purpose, we closely 
examine the raw NRG spectra for resonance widths as small as $10^{-4}\w_c$,
using reasonably {\it few} interleaved averaging $N_z=20$, much smaller that
the naively needed $10^4$ NRG runs. Fig.~\ref{fig:raw} shows that for frequencies
far from the resonance located at $\Delta$, the highest weight NRG states come in 
packets of $N_z$ peaks for a given Wilson shell. Surprisingly, the density of NRG eigenstates 
demonstrates a huge increase precisely at the peak value, so that information about 
the resonance width and height seems really encoded in the discrete results! 
The standard broadening $b=0.7/N_z=0.035$ is however way too large to benefit from 
this effect, and the corresponding smoothed spectra are indeed completely inaccurate. 
For definite comparison, we have plotted the analytical result obtained from the expansion 
at weak dissipation~\cite{Shnirman}:
\begin{eqnarray}
\label{eq:analytic}
C(\w) & = & -\frac{1}{\pi} \mr{Im} \frac{\w-\Sigma(\w)}{\w[\w-\Sigma(\w)]-\Delta^2}\\
\nonumber
\Sigma(\w) & = & \int \frac{\mr{d}\epsilon}{\pi}
\frac{J(\epsilon)}{\w-\epsilon+i0^+} =  \int_0^{\w_c}\!\!
\mr{d}\epsilon \, \frac{2\alpha \w_c^{1-s} \epsilon^s}{\w-\epsilon+i0^+}.
\end{eqnarray}

In order to judiciously exploit this unexpected finding, the straightforward 
idea we follow here is to adapt the broadening parameter $b$ in~(\ref{eq:broaden})
to the frequency dependence of the local density of NRG peaks. A very natural
approach adapted to the NRG logarithmic discretization is to extract $b(\w)$ from 
the logarithmic derivative of the integrated spectrum up to frequency $\w$:
\begin{equation}
b(\w) = \frac{b_0}{2}\left(\left[q+\frac{\mr{d}\log \int_0^\w
C}{\mr{d}\log\w}\right]^{-1}\!\!\!\!\!\!+\!\left[q+\frac{\mr{d}\log \int_{+\infty}^\w
C}{\mr{d}\log\w}\right]^{-1}\right)
\label{eq:newbroaden}
\end{equation}
where $q\simeq1$ is regularization parameter, whose precise value is not
sensitive to the final results, and $b_0$ sets the typical broadening at 
very low and very high frequencies (far from the atomic resonances). Note
indeed that we average in this expression two frequency sweeps from $\w=0$ and
$\w=+\infty$ respectively, in order to treat on an equal basis low and high 
frequency tails.
Because the actual NRG data is fully discrete~(\ref{eq:discrete}),
we extract $b(\w)$ recursively using equation~(\ref{eq:newbroaden}) on the broadened
NRG spectra. This procedure converges after few iterations to the results
displayed on Figs.~\ref{fig:raw} and~\ref{fig:compare_verysharp}. 
\begin{figure}[ht]
\includegraphics[width=8.5cm]{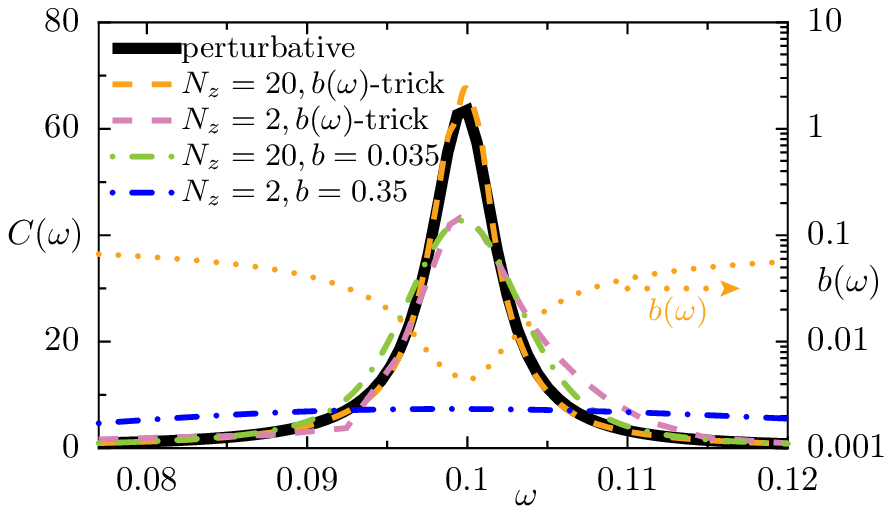}
\caption{(Color online) Similar comparison as done on Fig.~\ref{fig:compare_verysharp}, 
with stronger dissipation $\alpha=0.02$. Optimized broadenings with $b(\w)$-trick were
done with $N_z=2,20$ and $b_0=0.6,0.12$ respectively.}
\label{fig:compare_sharp}
\end{figure}
Given the small numerical effort devoted to generate these NRG data, the quality of 
final result is quite astonishing. The reason for this success is given on
Fig.~\ref{fig:compare_verysharp}, where the $b(\w)$ parameter takes values as 
small as $10^{-4}$ at the resonance, while it increases drastically far away from it, 
naturally cancelling out a great part of the NRG oscillations due to the discretization 
of the Wilson chain. We emphasize that the only free parameter is 
the typical low-frequency broadening $b_0$ in~(\ref{eq:newbroaden}), which is easily 
adjusted: we always find a range of $b_0$ values where the self-consistently broadened 
spectrum is relatively independent of the chosen $b_0$.
%(for too small or large $b_0$, the spectra are on the other hand clearly under or overbroadened).
More generally, one can also avoid the use of Ansatz~(\ref{eq:newbroaden}) by determining
$b(\w)$ from estimates of the local density of highest weight states in the raw NRG
data, giving similar results.
Large scale NRG simulations with $N_z\gg10$ and fixed $b$, as performed 
e.g. in~\cite{Zitko}, may thus benefit greatly from an adaptive broadening
procedure. We also note that 
the high frequency tails are also faithfully reproduced using the $b(\w)$-trick. 
Finally, in situations where relatively broader resonances are present, we show
that quantitative results can be obtained at lower computational cost, by
decreasing significantly the number of $z$-averaging even as low as $N_z<5$, see 
Fig.~\ref{fig:compare_sharp}. 

\begin{figure}[ht]
\includegraphics[width=8.7cm]{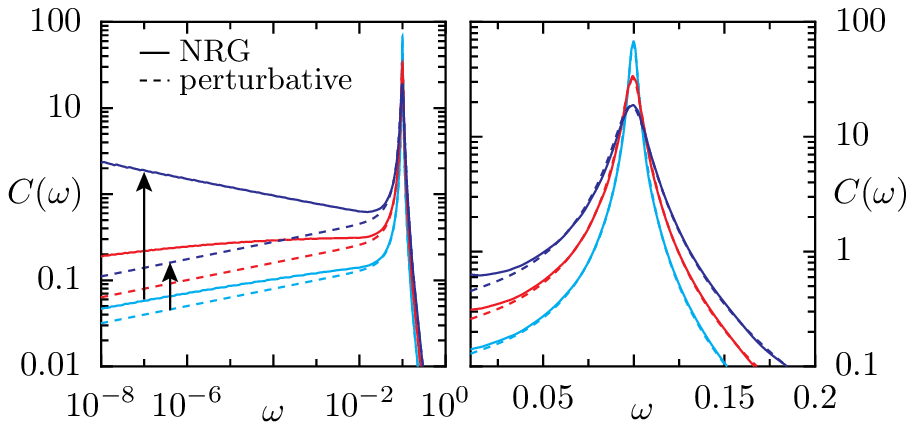}
\caption{(Color online) Comparison of optimized NRG spectra (solid lines) and 
perturbative calculations (dashed lines) with bath exponent $s=0.1$,
$\Delta=0.1\w_c$ and increasing values (arrow) of dissipation $\alpha=0.002,
0.004, 0.007$.}
%The right panel shows a close up on the resonance.} 

\label{fig:comparison_s0.1}
\end{figure}

\begin{figure}[ht!]
\includegraphics[width=8.7cm]{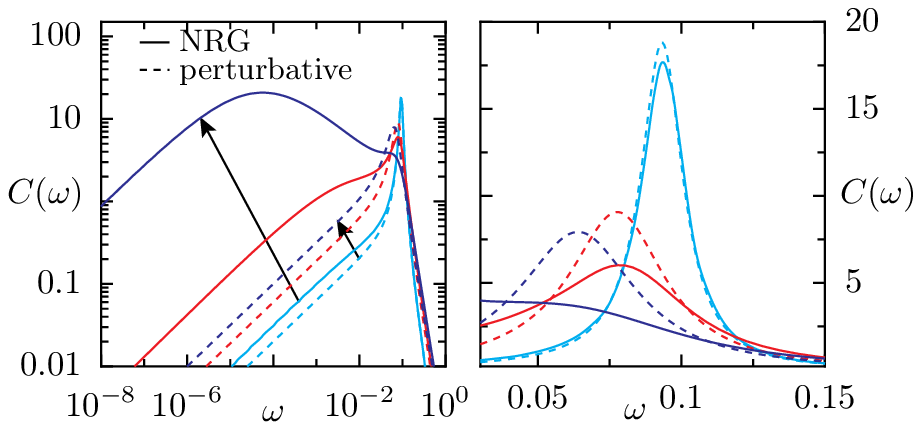}
\caption{(Color online) Similar plot as on Fig.~\ref{fig:comparison_s0.1},
with bath exponent $s=0.5$, $\Delta=0.1\w_c$ and $\alpha=0.02, 0.06, 0.10$.}
\label{fig:comparison_s0.5}
\end{figure}

In order to check that our broadening procedure is completely robust for the
whole range of parameters in the model~(\ref{eq:ham}), we investigate the effect
of increasing dissipation for several bath exponent values $s$. The quantitative 
comparison of the Rabi resonance to the analytical formula~(\ref{eq:analytic}) in 
its domain of validity, namely $\alpha\ll1$, gives us indeed confidence in the 
adaptative method.
This is proved on Fig.~\ref{fig:comparison_s0.1} and \ref{fig:comparison_s0.5}, 
which consider the same parameters as in Ref.~\onlinecite{Anders}, with important 
quantitative improvement. 
For the small value of $s=0.1$ taken on Fig.~\ref{fig:comparison_s0.1}, the perfect 
matching of the NRG data on the resonance to the lowest order calculation in 
$\alpha$ shows that the dissipation mechanism does not care for the low-energy 
behavior of the spin dynamics, even at the quantum critical
point~\cite{Tong,Anders} where the spin localizes for the values $\alpha_c=0.0071$ 
and $0.105$ with $s=0.1$ and $0.5$ respectively ($\Delta/\omega_c=0.1$ and 
$\Lambda=2$ here). Interestingly, the low-energy tails show significant deviations 
from the analytical result~(\ref{eq:analytic}) even far from the quantum phase 
transition, so that lowest-order calculations~\cite{Shnirman} do not apply for
the long-time dynamics.
Improved resummation of the perturbation theory to all orders in $\alpha$ 
will be considered in a forthcoming work~\cite{Florens}. For the intermediate value
$s=0.5$ shown on Fig.~\ref{fig:comparison_s0.5}, clear non-perturbative effects
on the resonance are seen inbetween the weak coupling regime and the quantum
critical point, so that perturbation theory in $\alpha$ breaks down.

To conclude, we have investigated the properties of discrete NRG spectra 
near atomic states in the spin boson model, and found that a bunching of the 
many-body levels occurs whenever sharp resonances are encountered. An 
adaptive broadening scheme was proposed, showing a drastic improvement in 
computation power for the calculation of accurate spectral functions 
over the whole energy range. This procedure will certainly allow to take further 
advantage of the potentialities of the NRG in a wide range of physical situations, 
from ab-initio calculations for magnetic impurities in metals~\cite{Costi3} to 
the difficult problem of simulating strongly correlated materials in the framework 
of the Dynamical Mean Field Theory~\cite{DMFT}.


\begin{thebibliography}{50}

\bibitem{Hewson} A. C. Hewson, {\it The Kondo Problem to Heavy
Fermions}, Cambridge University Press, Cambridge (1996).
\bibitem{Wilson} K.~G. Wilson, Rev. Mod. Phys. {\bf 47}, 773 (1975).
\bibitem{Krishnamurthy} H.~R. Krishna-murthy, J.~W. Wilkins and K.~G. Wilson,
Phys. Rev. B {\bf 21} 1003 (1980).
\bibitem{Bulla1} R. Bulla, T. Costi and T. Pruschke, Rev. Mod. Phys. {\bf 80},
395 (2008).
\bibitem{Costi1} T.~A. Costi, A.~C. Hewson and V. Zlatic, J. Phys.: Condens.
Matter {\bf 6}, 2519 (1994).
\bibitem{Tong} R. Bulla, H.~J. Lee, N.~H. Tong and M. Vojta, Phys. Rev. B 
{\bf 71}, 045122 (2005).
\bibitem{Costi2} T.~A. Costi, Phys. Rev. Lett. {\bf 85}, 1504 (2000).
\bibitem{Bulla2} R. Bulla, Phys. Rev. Lett. {\bf 83}, 136 (1999).
\bibitem{DMFT} A. Georges, G. Kotliar, W. Krauth and M. Rozenberg, Rev. Mod.
Phys. {\bf 68}, 13 (1996).
\bibitem{Sachdev} S. Sachdev, {\it Quantum Phase Transitions}, Cambridge
University Press, Cambridge (1999).
\bibitem{Vojta} M. Vojta, Phil. Mag. {\bf 86}, 1807 (2006).
\bibitem{Potok} R.~M. Potok, I.~G. Rau, H. Shtrikman, Y. Oreg, and D.
Goldhaber-Gordon,  Nature {\bf 446}, 167 (2007).
\bibitem{Roch} N. Roch, S. Florens, V. Bouchiat, W.
Wernsdorfer and F. Balestro, Nature {\bf 453}, 633 (2008).
\bibitem{Oliveira1} W.~C. Oliveira and L.~N. Oliveira, Phys. Rev. B {\bf 49},
11986 (1994).
\bibitem{Zitko} R. Zitko and T. Pruschke, Phys. Rev. B {\bf 79}, 085106 (2009).
\bibitem{Leggett} A. J. Leggett, S. Chakravarty, A. T. Dorsey, M. P. A. Fisher,
A. Garg and W. Zwerger, Rev. Mod. Phys. {\bf 59}, 1 (1987).
\bibitem{Anders} F.~B. Anders, R. Bulla and M. Vojta, Phys. Rev.
Lett. {\bf 98}, 210402 (2007).
\bibitem{Bulla3} R. Bulla, T. A. Costi, and D. Vollhardt,
\prb {\bf 64}, 045103 (2001).
\bibitem{Weichselbaum} A. Weichselbaum and J. von Delft, Phys. Rev. Lett. {\bf
99}, 076402 (2007); R. Peters, T. Pruschke and F.~B. Anders, Phys. Rev. B {\bf
74}, 245114 (2006).
%\bibitem{Karski} M. Karski, C. Raas and G.~S. Uhrig, Phys. Rev. B {\bf 77},
%075116 (2008).
\bibitem{Oliveira2} V.~L. Campo and L.~N. Oliveira, Phys. Rev. B {\bf 72}, 104432 (2005).
\bibitem{Weichselbaum2} A. Weichselbaum, F. Verstraete, U. Schollw\"{o}ck, J.~I.
Cirac and J. von~Delft, preprint {\tt arXiv:cond-mat/0504305}.
\bibitem{Shnirman} A. Shnirman and Y. Makhlin, Phys. Rev. Lett. {\bf 91}, 207204 (2003).
\bibitem{Florens} S. Florens, A. Freyn, R. Narayanan and D. Venturelli, {\it in
preparation}.
\bibitem{Costi3} T.~A. Costi {\it et al.}, Phys. Rev. Lett. {\bf 102}, 056802 (2009).
%\bibitem{Costi3} T.~A. Costi, L. Bergqvist, A. Weichselbaum, J. von Delft, T.
%Micklitz, A. Rosch, P. Mavropoulos, P.~H. Dederichs, F. Mallet, L. Saminadayar
%and C. Bauerle, preprint {\tt arXiv:0810.1771}.
\end{thebibliography}
\end{document}